\documentclass[a4paper,11pt]{article}

\usepackage{jcappub} % for details on the use of the package, please see the JINST-author-manual
\usepackage[T1]{fontenc} % if needed
\usepackage{graphicx}% Include figure files
\usepackage{dcolumn}% Align table columns on decimal point
\usepackage{bm}% bold math
\usepackage{amsmath}
\usepackage{amsfonts} % for kokuban-shotai
\usepackage{latexsym}
\usepackage{bbm}
\usepackage{color}
\usepackage{amssymb}
\usepackage{lipsum, babel}
\usepackage{amsthm}
\usepackage{epsf}
\usepackage{epsfig}
\usepackage{caption3}
\usepackage{appendix}
\usepackage{cases}
\usepackage{subfigure}
\usepackage{multirow}
\usepackage[inkscapelatex=false]{svg}
\usepackage{cleveref}
\newcommand{\Rmnum}[1]{\expandafter\@slowromancap\romannumeral #1@}
\makeatother

\makeatother

\newcommand{\calL}{{\cal L}}

\title{Estimating the gravitational wave background anisotropy: a Bayesian approach boosted by cross-correlation angular power spectrum}

\author[a]{Chi Tian,}
\author[a,1]{Ran Ding\note{Corresponding authors.},}
\author[b,1]{and Xiao-Xiao Kou}

\affiliation[a]{School of Physics and Optoelectronics Engineering, Anhui University, 111 Jiulong Road, Hefei, Anhui, 230601, China}
\affiliation[b]{School of Physics and Astronomy, University of Minnesota, Minneapolis, MN 55455, USA}

\emailAdd{ctian@ahu.edu.cn}
\emailAdd{dingran@mail.nankai.edu.cn}
\emailAdd{kou00016@umn.edu}

\abstract{
We introduce a new method designed for Bayesian inference of the angular power spectrum of the  Gravitational Wave Background (GWB) anisotropy. This scheme works with detector data and can optionally incorporate the cross-correlations between the GWB  anisotropy and other cosmological tracers, enhancing the significance of Bayesian inference. We employ the realistic LISA response and noise model to demonstrate the validity of this approach. The findings indicate that, for the injected cosmological GWB signal, without considering any cross-correlations, the 4-year LISA data is insufficient to achieve a significant detection of multipoles. However, if the anisotropies in the GWB are strongly correlated with the Cosmic Microwave Background (CMB), the 4-year data can provide unbiased estimates of the quadrupole moment ($\ell = 2$).  This reconstruction process is generic and not restricted to any specific detector, offering a new framework for extracting anisotropies in the GWB data from various current and future gravitational wave observatories.}

\begin{document}
\maketitle
\flushbottom

\section{Introduction} 
Our ability to detect gravitational waves (GWs) will undoubtedly reach an unprecedented level in the next decade, with detection ranges spanning from nanohertz to several hundred hertz and significantly enhanced precision. 
This substantial advancement is likely to help us fully decipher a new background signal in the Universe — the gravitational wave background (GWB),  whose preliminary clues have been revealed by recent Pulsar Timing Array (PTA) experiments~\citep{NANOGrav:2023gor,Xu:2023wog,EPTA:2023fyk,Reardon:2023gzh, Miles:2024seg}. 
This diffused GW signal, resulting from  superpositions of numerous unresolved sources, can be classified into two categories: the astrophysical gravitational wave background (AGWB) and the cosmological gravitational wave background (CGWB). The former results from superpositions of unresolved galactic and extra-galactic sources, mostly related to the dynamics of compact objects \citep{Ferrari:1998jf,Ferrari:1998ut,Schneider:2000sg,Farmer:2003pa,Regimbau:2011rp,Zhu:2011bd}. The latter, originating from early Universe processes, is closely related to new physics, such as inflation, reheating/preheating, primordial black holes,  topological defects, or cosmological first-order phase transitions \citep{Vachaspati:1984gt,Kosowsky:1992rz,Guzzetti:2016mkm,Caprini:2018mtu,Aggarwal:2020olq}.

Similar to the CMB and other cosmological probes, the GWB also carries anisotropies, which can originate from either source distributions or propagation processes. Anisotropies arising from sources include those generated by inhomogeneous distributions of compact objects \citep{Alba:2015cms}, in the case of the AGWB, or those resulting from large scale perturbations introduced by domain walls in the early universe \citep{Liu:2020mru}, as in the CGWB. On the other hand, the propagation process may lead to anisotropies due to  various gravitational effects, analogous to the CMB photons \citep{Contaldi:2016koz}. Considerable efforts have been made in recent years to model the angular power spectrum of the GWB anisotropy \citep{Geller:2018mwu, Bartolo:2019oiq,Bartolo:2019yeu,ValbusaDallArmi:2020ifo,Li:2021iva,Schulze:2023ich, Cui:2023dlo,Wang:2023ost, Li:2023qua, ValbusaDallArmi:2023ydl, Bethke:2013aba, Cusin:2017fwz,Cusin:2018rsq,Jenkins:2018kxc,Cusin:2019jpv,Bertacca:2019fnt,Bellomo:2021mer,Sato-Polito:2023spo,Pol:2022sjn,Malhotra:2022ply,Yu:2023jrs, Cai:2024dya} and to assess its detectability in the operating and proposed GW observations through estimating the signal-to-noise ratio (SNR) \citep{Alonso:2020mva, Alonso:2020rar, Ricciardone:2021kel,  Capurri:2021prz, Jenkins:2018uac, Zhao:2024yau, Capurri:2022lze}. 
% These investigations have primarily focused on signal-to-noise ratio (SNR) estimations or frequentist sky map reconstructions \citep{Renzini:2018vkx, Renzini:2019vmt, Banagiri_2021, LISACosmologyWorkingGroup:2022kbp, Chung:2023rpq, Tsukada:2022nsu,Gair:2014rwa, Li:2024lvt}. 
Meanwhile, to quantitatively extract anisotropic information from the data, both Bayesian and frequentist frameworks for constructing GWB sky maps have been developed~\citep{Renzini:2018vkx, Renzini:2019vmt, Banagiri_2021,LISACosmologyWorkingGroup:2022kbp,Chung:2023rpq,Tsukada:2022nsu,Gair:2014rwa,Li:2024lvt,NANOGrav:2023tcn}.  However, existing Bayesian map-making techniques, which rely on constructing the entire sky map using the spherical harmonic approach, often encounter challenges such as slow convergence (see discussions in \cite{NANOGrav:2023tcn}). Given that many physical models predict the power spectrum, a Bayesian framework designed to extract the angular spectrum directly from detector data could be more suitable and convenient for exploring these models. Moreover, as predicted by many theoretical studies \citep{Adshead:2020bji,Ricciardone:2021kel, Ding:2023xeg, Schulze:2023ich, Yang:2023eqi}, regardless of their astrophysical or cosmological  origins, the anisotropies in the GWB generally exhibit significant cross-correlations with other existing cosmological tracers, such as the CMB, the CMB lensing, or large-scale structures. The detection sensitivity to the anisotropies would undoubtedly benefit from these potential cross-correlations, as suggested by \citep{Capurri:2021prz, Alonso:2020mva, Ricciardone:2021kel}, through estimating the enhanced SNR. Nevertheless, a consistent methodology for integrating these cross-correlations into the data analysis framework, facilitating the extraction of the angular power spectrum  of the GWB anisotropy, has not been established. 

In this study, we focus on a novel data analysis scheme designed to extract anisotropies in the form of the angular power spectrum from data of GW detectors without constructing any sky map. This framework optionally accounts for the potential cross-correlations between the GWB anisotropy and other cosmological tracers. We demonstrate the validity of this approach by applying it to LISA \citep{LISA:2017pwj, LISA:2022kgy,LISA:2024hlh} mock data and discuss its broader applications for future GW observations.

\section{Anisotropic GWB intensity and detector data}
Assuming the GWs to be Gaussian, stationary and unpolarized, the metric perturbation $h_{p}(f, \hat{\mathbf{n}})$ in Fourier space can be characterized by its quadratic expectation values as
\begin{align}
\left\langle h_p(f, \hat{\mathbf{n}}) h_{p^{\prime}}^*\left(f^{\prime}, \hat{\mathbf{n}}^{\prime}\right)\right\rangle \equiv \delta(f-f') \frac{\delta^2\left(\hat{\mathbf{n}},\hat{\mathbf{n}}^{\prime}\right)}{4 \pi} \mathrm{W}_{p p^{\prime}}(f, \hat{\mathbf{n}}),
\end{align}where $p=\{+, \times\}$ labels the polarization states, $\hat{\mathbf{n}}$ is the direction of propagation in the sky and the matrix $W$ is related to the gravitational Stokes parameters $Q$, $U$, and $V$.
Under the unpolarized condition, $Q=U=V=0$, one has $\mathrm{W}_{++, \times \times}(f, \hat{\mathbf{n}}) = I(f, \hat{\mathbf{n}})$, where $ I(f, \hat{\mathbf{n}})$ represents the GW intensity.

On the other hand, when considering the GWB, it's a common practice to quantifying it through the fractional energy density $\Omega_{rm GW}$, whose spectrum can be separated into a background part (monopole) and a linear fluctuation. It can be proven \citep{Smith:2019wny, LISACosmologyWorkingGroup:2022kbp} the relationship between the anisotropic GW intensity and the fractional energy density, using the spherical harmonic coefficients, can be expressed by (see Appendix \ref{app:a1} for the detailed derivation)
\begin{align}\label{eq:I00}
I_{0 0}(f) \equiv \frac{3 H_0^2}{4\pi^2}\frac{\overline{\Omega}_{\mathrm{GW}}(f)}{f^3},
\end{align}and 
\begin{align}
\label{eq:Kf}
I_{\ell m}(f) &= \frac{3 H_0^2}{4\pi^2}\frac{\overline{\Omega}_{\mathrm{GW}}(f)}{\sqrt{4\pi} f^3} \left(4-\frac{\partial \ln \overline{\Omega}_{\mathrm{GW}}\left(f \right)}{\partial \ln f}\right) a^{\rm GW}_{\ell m}\nonumber\\
&\equiv K_f a^{\rm GW}_{\ell m},
\end{align}
where we have assumed that $K_f$ has encapsulated all the frequency dependencies, leaving $a_{\ell m}^{\mathrm{GW}}$s frequency-independent \cite{Thrane:2009fp,Romano:2016dpx}. Our aim is to extract the power spectrum of $a_{\ell m}^{\mathrm{GW}}$, denoted as $C_{\ell}^{\mathrm{GW}}$, from the GWB data.

Our data analysis is based on the short-time Fourier transform (STFT) data of a GW detector A,  denoted as  $d^A_t(f)$, which is obtained by applying Fourier transforms to the time-series signal over a specific time segment $T_{\mathrm{seg}}$, around a given time frame $t$ as
\begin{align}
    d^A_t(f) = \int_{t-T_{\mathrm{seg}}/2}^{t+T_{\mathrm{seg}}/2} d t' e^{-2 i \pi f t'} d^A(t').
\end{align}
Note that $T_{\rm seg}$ also determines the width of frequency bins $\Delta f = 1/T_{\rm seg}$ and $d^A_t(f)$ can be estimated through
\begin{align}
d^A_t(f)=\int d^2  \Omega_{\hat{\mathbf{n}}} \sum_p R^A_{t, p}(f, \hat{\mathbf{n}}) h_p(f, \hat{\mathbf{n}})+ n^A_{t, f}\,,
\label{eq:22}
\end{align}where $R^A_{t, p}(f, \hat{\mathbf{n}})$ and $n^A_{t, f}$  are the response function and noise for the detector $A$.  $d^A_t(f)$ has a zero mean, and its ensemble average, evaluated at equal time frame $t$ (and is otherwise zero), can be estimated by
\begin{align}
\label{eq:cov_d}
\left\langle d^A_t(f) d^{B*}_{t}\left(f'\right)\right\rangle  &= \delta(f -f')\left[\sum_{\ell=0}^{\infty} \sum_{m=-\ell}^\ell \gamma^{AB}_{\ell m}(t, f) I_{\ell m} (f) + N^{AB}_{t, f}\right] \nonumber \\
&\equiv    \frac{\delta^K_{f f'}}{\Delta f} \langle D^{AB}_{t, f}\rangle\,, 
\end{align}
where the Dirac delta function $\delta(f-f')$ is replaced by Kronecker delta $\delta^K$ through $\delta(f-f') = \delta^K_{ff'}/\Delta f$, and 
$ \langle n^A_{t, f} n^{B*}_{t, f'}\rangle \equiv \delta(f-f') N^{AB}_{t, f}$ represents the noise power spectral density (PSD) for the detector pair $AB$,  and $\gamma^{AB}_{\ell m}(t, f)$ are spherical harmonic coefficients of the response function $\gamma^{AB}_{t, f}( \hat{\mathbf{n}}) $ for this detector pair, defined by 
\begin{align}
\label{eq:res_g}
\gamma^{AB}_{t,f}(\hat{\mathbf{n}}) = \Delta \Omega \sum_p R^A_{t, p}(f, \hat{\mathbf{n}}) R^{B*}_{t, p}(f, \hat{\mathbf{n}})\,.
\end{align}
In similar manner, the covariance of $D^{AB}_{t, f}$ can be written as
\begin{align}
\label{eq:cov_D}
C_{t,f}^{ABCD} \equiv \langle D^{AB}_{t, f}, D^{CD*}_{t, f}\rangle 
\approx \left(\gamma^{AC}_{00}(t, f) I_{00} (f) + N^{AC}_{t,f}\right)
\left(\gamma^{DB}_{00}(t, f) I_{00} (f) + N^{DB}_{t, f}\right)\,,
\end{align}where the approximation adopted is based on the fact that the contribution from multipoles are subdominant compared to the those from monopole. As a consequence, the information of anisotropies is exclusively encoded in $\langle D^{AB}_{t, f}\rangle$. 
Note that while a diagonalized covariance matrix is often employed, such as the  time-delay interferometry (TDI) channels frequently used in the LISA literature, this formalism is more general. It does not require a diagonalized covariance matrix and inherently accounts for cross-detector correlations. The conversion between our scheme and the TDI channels can be found in \cite{Alonso:2020rar}. 

\section{Likelihood with cross-correlations}
Our aim is to perform Bayesian estimations on the angular power spectrum of the GWB anisotropy using the detector dataset $ D^{\{AB\}}_{\{t, f\}}$.  For this purpose,  a practical likelihood function is necessary.  When a set of $I_{{\{\ell m\}}, f}$ is given, the likelihood for $ D^{\{AB\}}_{t, f}$, which represents data for various detector's combinations at a specific time and frequency, can be simply written as
\begin{align}
\label{eq:p_D_I}
& \calL\left(D_{t, f}\mid I_{\{\ell m\}, f}\right) \nonumber.\\
=& \prod_{ AB, CD
}\frac{1}{\big|\pi C_{t,f}^{ABCD}\big|}\exp{\left\{-\Big(D^{AB}_{t, f} - \langle D^{AB}_{t, f}\rangle\Big)^\dagger  \left(C^{ABCD}_{t, f}\right)^{-1}\Big(D^{CD}_{t, f} - \langle D^{CD}_{t, f}\rangle\Big)\right\}}\,.
\end{align}
Note that this likelihood is valid only when $D^{\{AB\}}_{\{t, f\}}$ are Gaussian, which is approximately the case when considering $D^{\{AB\}}_{\{t, f\}}$ as the average of multiple samples. However, for a given angular power spectrum $C_{\{\ell\}}^{\mathrm{GW}}$, there are in principle an infinite number of realizations in terms of $I_{\{\ell m\}}$,  making it challenging to write down a likelihood function that directly links the detector data with a model predicted $C_{\{\ell\}}^{\mathrm{GW}}$. Therefore, we must remain agnostic regarding specific realizations.

In addition, the GWB anisotropy can be correlated with another known cosmological tracer $\mathrm{Y}$. When neglecting potential non-Gaussianities, analogous to studies investigating the cross-correlation between the CMB and other probes (i.e. \cite{PhysRevD.71.063005, PhysRevD.78.043520,Hu:2004yd}), we assume that the maps of the GWB and the tracer $\mathrm{Y}$ can be described by multivariate Gaussian variables. Their covariance is determined by their auto angular spectra $C_{\ell}^{\mathrm{GW}}$, $C_{\ell}^{\mathrm{Y}}$ and the cross spectrum $C_{\ell}^{\mathrm{GW}\times\mathrm{Y}}$. Consequently, the conditional probability for the GWB can be derived with the knowledge of  ${\mathrm{Y}}$, whose measured spherical harmonics coefficients are $a_{\{lm\}}^{\mathrm{Y}}$ with negligible uncertainties compared to the GWB measurements.  It can be proven (see Appendix \ref{app:a2} for the detailed derivation) that for a given condition characterized  by the combination of $\left(a^{\mathrm{Y}}_{\{\ell m \}}, C_{\{\ell\}}^{\mathrm{GW}}, C_{\{\ell\}}^{\mathrm{GW}\times \mathrm{Y}}\right)$, the conditional distribution of $I_{\{\ell m\}}$ at a given frequency, denoted as $\calL\left(I_{\{\ell m \}, f}\mid  \mathrm{GW}\times \mathrm{Y}\right)$, obeys 
\begin{align}
\label{eq:p_Ilm_Y}
 \calL\left(I_{\{\ell m \},f}\mid  \mathrm{GW}\times \mathrm{Y}\right)   
= \prod_{\substack{\ell m\\\ell \geq 1}} \frac{1}{\pi C_{\ell}^{\mathrm{GW}\mid {\mathrm{Y}}}} \exp{\left(-\frac{\Big|I_{\ell m, f} - K_f \mu^{\mathrm{GW}\mid Y}_{\ell m} \Big|^2}{K_f^2 C_{\ell}^{\mathrm{GW}\mid {\mathrm{Y}}}}\right)}\, ,
\end{align}
where we have defined the mean and the auto-correlation of this GWB anisotropy as
\begin{align}
\mu^{\mathrm{GW}\mid {\mathrm{Y}}}_{\ell m} &\equiv \frac{C_{\ell}^{\mathrm{GW} \times {\mathrm{Y}}}}{C_{\ell}^{{\mathrm{Y}}}} a^{\mathrm{Y}}_{\ell m} \label{eq:mu} \, ,\\ 
C_{\ell}^{\mathrm{GW}\mid {\mathrm{Y}}} & \equiv C_{\ell}^\mathrm{GW} -\frac{\left(C_{\ell}^{\mathrm{GW} \times {\mathrm{Y}}}\right)^2}{C_{\ell}^{\mathrm{Y}}}  \label{eq:Cl_cross} \\\nonumber 
&= C_{\ell}^\mathrm{GW}\left[1-\left(r^{\mathrm{GW}\times\mathrm{Y}}\right)^2\right]\, ,
\end{align}for $\ell \geq1$, and the relative cross-correlation $r^{\mathrm{GW}\times\mathrm{Y}}$ is defined by $r^{\mathrm{GW}\times\mathrm{Y}} \equiv C_{\ell}^{\mathrm{GW} \times {\mathrm{Y}}}/\sqrt{C_{\ell}^\mathrm{GW}C_{\ell}^{\mathrm{Y}}}$. Note that since the intensity monopole $I_{00}$ is related to the GW fractional energy, as in eq.~\eqref{eq:I00},  the product in eq.~\eqref{eq:p_Ilm_Y} does not include the $\ell = 0$ term.  
For the sake of conciseness and computational convenience, we introduce the following compact notations. We use Latin letters, such as \textit{a} and \textit{b}, to compress the detector combinations, such as $AB$, into a single index. Additionally, we use Greek letters $\mu$, $\nu$, etc., to denote combinations of $\ell m$ s. To be specific, we have:
\begin{align}
&a=\{AB\}= \underbrace{\{1,1\},\{1,2\},\cdots,\{1,n\},\cdots,\overbrace{\{n,1\},\cdots,\{n,n\}}^{n\,{\rm terms}}}_{n\times n\,{\rm terms}}\,, \nonumber\\
&\mu = \{\ell m \} = \underbrace{\{1,-1\},\{1,0\},\{1,1\},\cdots,\overbrace{\{\ell_{\max},-\ell_{\max}\},\cdots,\{\ell_{\max},\ell_{\max}\}}^{(2\ell_{\max}+1)\,{\rm terms}}}_{\ell_{\max}\left(\ell_{\max}+2\right)\,{\rm terms}}\,, \nonumber\\
& \Gamma_{a 0} \equiv \gamma^{\{AB\}}_{00}, \quad \Gamma_{a \mu} \equiv \gamma^{\{AB\}}_{\{\ell m \}}, \quad  X_0 = \overline{X}_0 \equiv I_{00}, \quad X_{\mu} \equiv I_{\{\ell m \}}, \quad \overline{X}_{\mu} \equiv K_f\mu^{\mathrm{GW}\mid {\mathrm{Y}}}_{\{\ell m \}}, \nonumber\\
& D_a \equiv D_{tf}^{AB},\quad N_a \equiv N_f^{AB},\quad S_a \equiv \langle D_{tf}^{AB}\rangle=\Gamma_{a 0} X_0 + \Gamma_{a \mu} X_{\mu} + N_a, \quad (C_D)_{ab} \equiv C_{t, f}^{ABCD} , \nonumber\\
& (C_I)_{\mu\nu}\equiv K^2_f {\rm diag}\Big(\underbrace{ C_{1}^{\mathrm{GW}\mid {\mathrm{Y}}}, C_{1}^{\mathrm{GW}\mid {\mathrm{Y}}}, C_{1}^{\mathrm{GW}\mid {\mathrm{Y}}},\cdots, \overbrace{C_{\ell_{\max}}^{\mathrm{GW}\mid {\mathrm{Y}}},\,\cdots,\, C_{\ell_{\max}}^{\mathrm{GW}\mid {\mathrm{Y}}}}^{(2\ell_{\max}+1)\,{\rm terms}}}_{\ell_{\max}\left(\ell_{\max}+2\right)\,{\rm terms}}\Big)\,,
\end{align}
Following this convention, the likelihood functions in eqs.~(\ref{eq:p_D_I}) and (\ref{eq:p_Ilm_Y}) are reformulated as
\begin{align}
& \calL\left(D_{t, f}\mid I_{\{\ell m \}, f}\right) =  \frac{1}{\big|\pi C_D\big|}\exp \Big\{-\Big(D_a- S_a\Big)^\dagger (C_D)^{-1}_{ab}\Big(D_b- S_b\Big)\Big\}\,, \label{eq:p_D_I2}\\
& \calL\left(I_{\{\ell m\}, f}\mid  \mathrm{GW}\times \mathrm{Y}\right) = \frac{1}{\big|\pi C_I\big|}\exp \Big\{-\Big(X_\mu-\overline{X}_\mu \Big)^\dagger (C_I)_{\mu\nu}^{-1} \Big(X_\nu-\overline{X}_\nu \Big)\Big\} \,\label{eq:p_Ilm_Y2}\,.
\end{align}

To finally derive the likelihood for the entire data set $D_{\{t, f\}}$ while remaining agnostic to the specific realization, we marginalize over  all $I_{\ell m}$s and sum over contributions from all time frames and frequency bins by computing
\begin{align}
\calL\Big(D_{\{t, f\}} \Big| \mathrm{GW} \times \mathrm{Y}\Big)  & \propto \prod_{t,f}\int \mathcal{D} I_{\substack{\ell m, f\\\ell \geq 1}}    \calL\left(D_{t,f}\Big| I_{\{\ell m \},f}\right)
 \calL\left(I_{\{\ell m \},f}\mid   \mathrm{GW} \times \mathrm{Y} \right)   \nonumber\\
=& \prod_{t, f} \frac{1}{\big|\pi C_D\big|\big|\pi C_I\big|}\int \Big[\prod_{\ell=1}^{\ell_{\max}} \prod_{m=-\ell}^{\ell} d^2 X_{\rho}\Big] \exp \Big\{-\Big(D_a -\overline{X}_0\Gamma_{0 a} -X_{\mu}\Gamma_{\mu a} - N_a\Big)^*(C_D)^{-1}_{ab} \nonumber\\
\quad &\Big(D_b-\Gamma_{b 0} \overline{X}_0-\Gamma_{b \nu} X_{\nu} - N_b\Big)-\Big(X_\mu-\overline{X}_\mu \Big)^\dagger (C_I)_{\mu\nu}^{-1}\Big(X_\nu-\overline{X}_\nu \Big)\Big\} \,,
\label{eq:p_CL_D}
\end{align}
where $d^2 X_{\rho}=d{\rm Re}X_{\rho} d{\rm Im} X_{\rho}$, and the dimension of the integral is $d=\ell_{\max}\left(\ell_{\max}+2\right)$. By plugging in the expression from  \cref{eq:cov_d,eq:cov_D,eq:p_D_I2,eq:p_Ilm_Y2}, and after some algebra, we finally arrive at a very compact form of  the likelihood function (see Appendix \ref{app:a3} for detail derivations):
\begin{align}
\label{eq:likl}
&\calL\left(D_{\{t,f\}} \mid \mathrm{GW}\times \mathrm{Y} \right) \propto \prod_{t, f}{\big|\pi C_{a b}\big|^{-1}}\exp \left(-J^{\dagger}_a C^{-1}_{a b} J_b\right)\,,\\
\label{eq:likl_2}
& C_{a b} = (C_D)_{ab} + \Gamma_{a\mu} (C_I)_{\mu\nu} \Gamma_{\nu b}^* \,, \\
\label{eq:likl_3}
& J_{a} = D_{a} - \Gamma_{a \mu} \overline{Z}_{\mu} - N_{a}\,. 
\end{align}
Note that all quantities in eq.~\eqref{eq:likl} depend on the  time and frequencies. we have also defined $\overline{Z}_{\mu} \equiv \mu^{\mathrm{GW}\mid {\mathrm{Y}}}_{\ell m}$  for $\ell, \mu \geq 1$ and $\overline{Z}_{0} \equiv I_{00}$ (see eq.~\eqref{eq:I00}) for $\ell,\mu= 0$. 

Eqs.~\eqref{eq:likl} to \eqref{eq:likl_3} consist the main results of this study. It is also important to know that this likelihood is generic, being applicable to models that are not considering any cross-correlations. In such case,  one has $C_{\ell}^{\mathrm{GW} \times {\mathrm{Y}}}=0$, and the mean and the covariance of harmonic coefficients  are reduced to $\overline{Z}_{\mu} = K_f\mu^{\mathrm{GW}\mid {\mathrm{Y}}}_{\ell m }=0$ for $\mu \geq 1$ and $C_{\ell}^{\mathrm{GW}\mid {\mathrm{Y}}}  = C_{\ell}^\mathrm{GW}$, producing a likelihood function based on the GW angular power spectrum without considering any cross-correlations, similarly as the results in \cite{vanHaasteren:2012hj, Gair:2014rwa}.

\section{Bayesian inference of LISA mock data}
\begin{figure}[htp]
    \centering
    \includegraphics[width=0.6\linewidth]{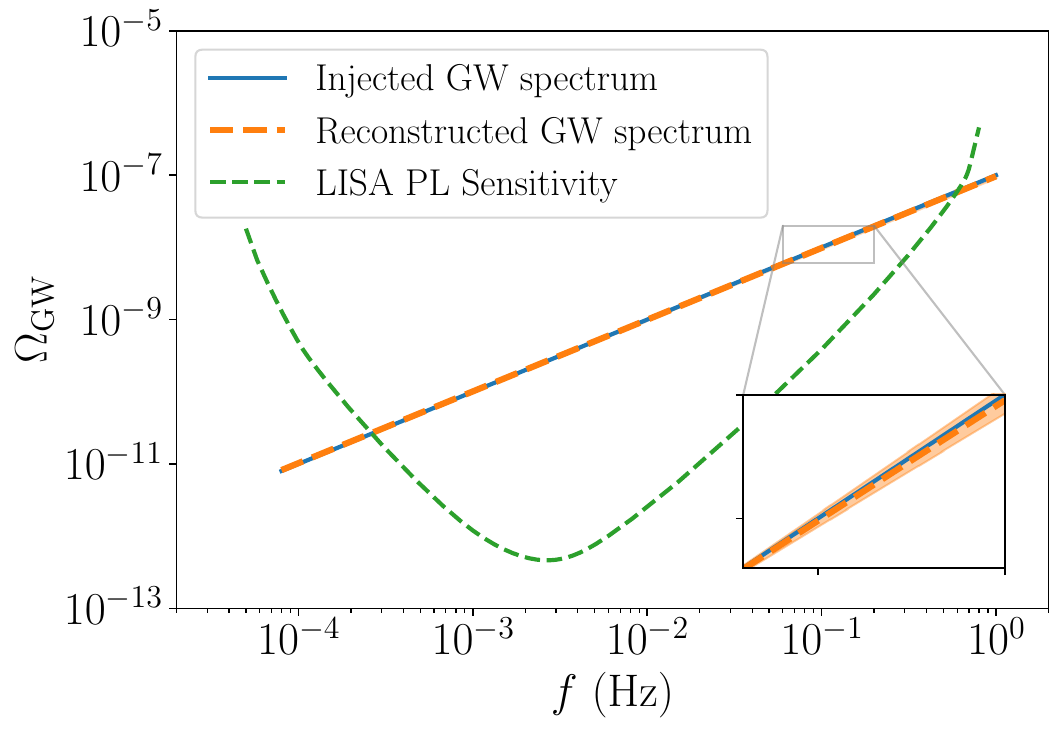}
    \includegraphics[width=0.6\linewidth]{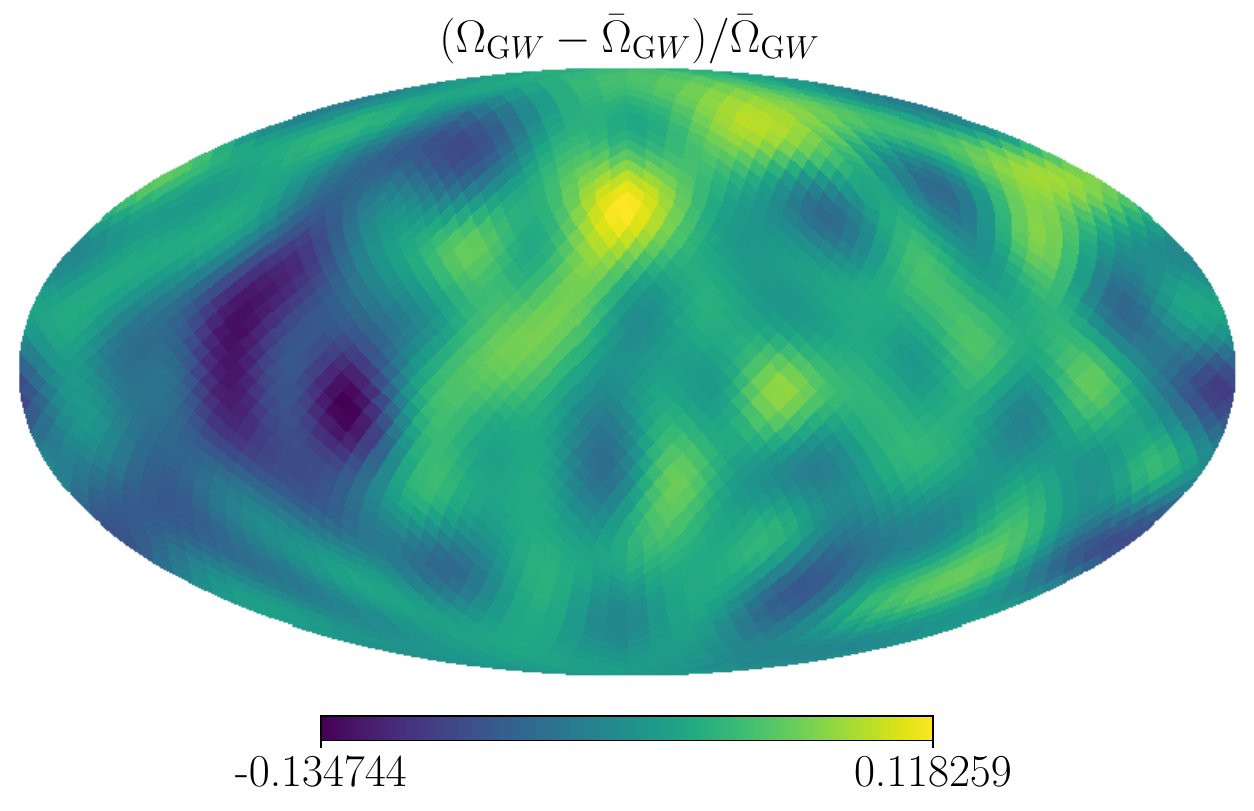}
    \caption{Upper Panel: Injected and reconstructed GWB spectrum. The $1\sigma$ uncertainty of the reconstruction is represented by the nearly invisible shaded region;
    Lower Panel: The sky map of the fluctuation of injected GWB ($(\Omega_{\mathrm GW}-\bar{\Omega}_{\mathrm GW})/\bar{\Omega}_{\mathrm GW}$), whose $\log_{10} \widetilde{C}_{\ell}^{\mathrm{GW}} = -2.4$. Locations of map pixels are following the \texttt{HealPix}\citep{Gorski:2004by} convention with $N_{\mathrm{side}}=16$.}
    \label{fig:inj}
\end{figure}

For demonstration purposes, we inject anisotropic GWB signals to LISA and attempt to estimate the injected anisotropic level.  The injected signal has a power-law spectrum $\overline{\Omega}_{\mathrm {GW}}(f) = A (f / f_{\mathrm{ref}})^{\alpha}$, where we  have chosen $A = 10^{-10}$, $f_{\mathrm{ref}}=10^{-3}\; \mathrm{Hz}$ and $\alpha = 1$, as depicted in the upper panel of Fig.~\ref{fig:inj}. The LISA power-law (PL) sensitivity curve (SNR=3 with 4-year observations), representing the detector's ability to measure a GWB spectrum in power-law form, is also presented in the same figure.
It is a relatively strong signal compared to the LISA PL sensitivity and the predicted spectrum of compact binaries \citep{Ferrari:1998jf,Ferrari:1998ut,Schneider:2000sg,Farmer:2003pa,Regimbau:2011rp}. Therefore, this mock signal can be considered as a GWB from new physics, which may arise from first order phase transitions or inflationary models (see \cite{Caprini:2024hue, Bartolo:2016ami} for comprehensive reviews) . 

The anisotropies in the signal are generated based on a scale invariant angular spectrum of $a_{\ell m}^{\mathrm{GW}}$s, implying that $\widetilde{C}_{\ell}^{\mathrm{GW}} \equiv \ell  (\ell + 1) C_{\ell}^{\mathrm{GW}}  = \mathrm{Const}$. This is a typical approximated prediction of many cosmological models at low $\ell$s \citep{Ricciardone:2021kel, Schulze:2023ich, Ding:2023xeg}, capturing propagation effects such as Sachs-Wolfe or Integrated Sachs-Wolfe contributions, along with anisotropies originating from sources. Our primary task is to estimate the $\widetilde{C}_{\ell}^{\mathrm{GW}} $ from the mock data. To this end, we generate a data set $\{{d^t_A(f)}\}$ following the covariance given by eq.~\eqref{eq:cov_d} in the frequency range $f\sim[0.0001, 0.2]\,\mathrm{Hz}$ with $\Delta f =10^{-6}\, \mathrm{Hz}$. This frequency spacing corresponds to an observation segment $T_{\mathrm{seg}}\sim 11 \, \mathrm{days}$, generating $25$ different observation segments in a year, corresponding to a $75\%$ observation efficiency. Each segment has a different antenna pattern due to the satellite array's  varying orientations  over a year. We compute the LISA response and the noise PSD using the public code \texttt{schNell} \citep{Alonso:2020rar}, which estimates the LISA time-dependent response function based on spacecraft positions over time, as derived in \cite{Rubbo:2003ap}, while assuming time-independent noise \citep{Smith:2019wny}.
Once the fiducial data is injected, we first attempt to recover the monopole spectrum of the signal (see Fig.~\ref{fig:inj}). This can be simply done by employing eq.~\eqref{eq:likl} and setting $C^I_{\mu \nu}$ to zero.  The nearly invisible shaded region indicates that the  uncertainties in the reconstructed monopole signals are negligible.  The small uncertainties suggest that the contributions from the multipoles in the likelihood can be safely ignored when inferring the monopole spectrum. We therefore fix the reconstructed monopole parameters during the subsequent inferences of the angular power spectrum.

\begin{figure}[htp]
    \centering
    \includegraphics[width=0.65\linewidth]{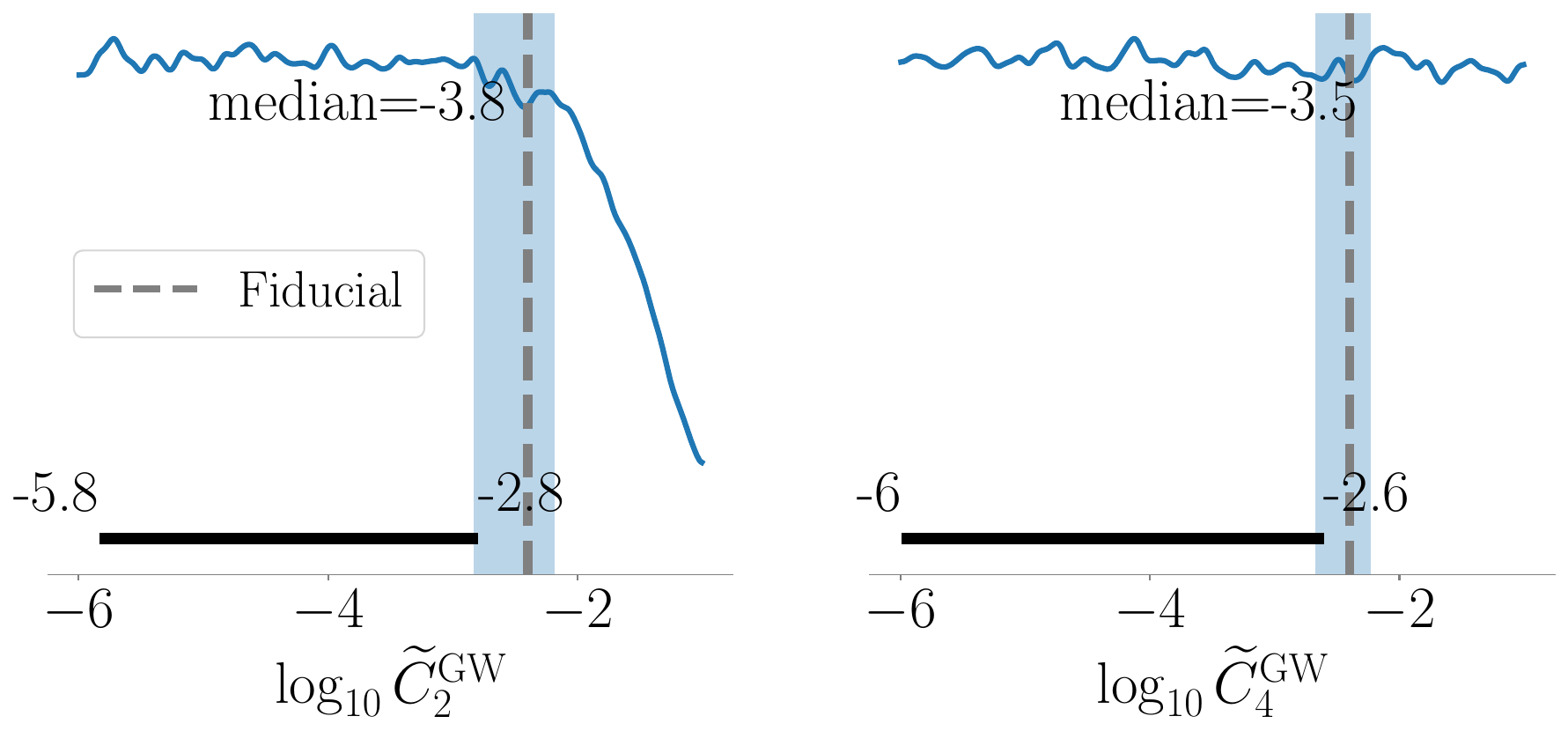}
    \includegraphics[width=0.65\linewidth]{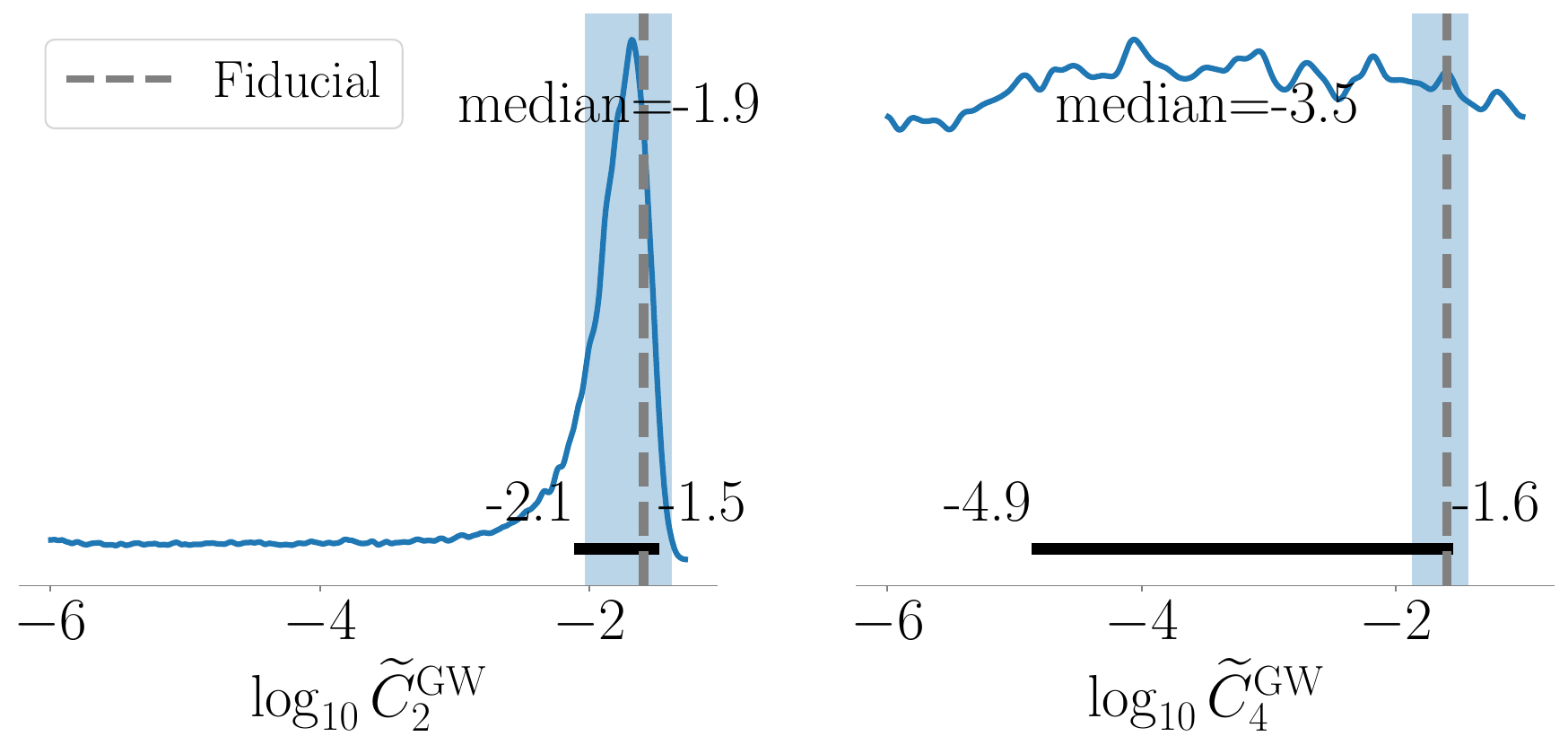}
    \caption{Upper Panel: Reconstructed multipoles $\widetilde{C}_{2}^{{\mathrm{GW}} }$ and $\widetilde{C}_{4}^{\mathrm{GW} }$ based on the LISA 4-year data. Fiducial $\log_{10} \widetilde{C}_{\ell}^{\mathrm{GW}} = -2.4$. Lower Panel: Reconstructed multipoles based on the LISA 80-year data. Fiducial $\log_{10} \widetilde{C}_{\ell}^{\mathrm{GW}} = -1.6$. The black bars on the bottom mark the $1\sigma$ region, and the shaded regions represent the cosmic variance.}
    \label{fig:4y_mult}
\end{figure}

We then employ the full likelihood in eq.~\eqref{eq:likl} to estimate the magnitude of the lowest several multipoles. To avoid potential numerical instabilities in the matrix inverses, in our calculation, we inverse matrices by expanding the term $\Gamma_{a\mu} C^I_{\mu\nu} \Gamma_{\nu b}^* $ in eq.~\eqref{eq:likl_2} to the 2nd order.  The resulting posteriors are presented in Fig. \ref{fig:4y_mult}. It is apparent that, with the injected anisotropic level in the cosmological GWB, the LISA 4-year mission lacks sufficient sensitivity to recover the magnitude of multipoles. This finding is consistent with previous studies~\citep{Alonso:2020mva, Capurri:2021prz, Capurri:2022lze} that are based on the SNR analysis. Note that we only present the results for even multipoles, as the even parity of the LISA antenna patten results in a weak SNR in the odd multipoles. We also find that unrealistically extending the LISA mission to 80 years and increasing the anisotropic level to $\log_{10} \widetilde{C}_{\ell}^{\mathrm{GW}} = -1.6$ will result in a restrictive posterior for the quadrupole moment $\widetilde{C}_2^{\mathrm{GW}}$, whose $1\sigma$ region is comparable to the cosmic variance $\sigma_{\widetilde{C}_{\ell}^{\mathrm{GW}}}=\sqrt{2/(2 \ell+1)}\widetilde{C}_{\ell}^{\mathrm{GW}}$ \citep{Bernardo:2022xzl}, thereby confirming the validity of our scheme.  

\begin{figure}[htp]
    \centering
    \includegraphics[width=0.65\linewidth]{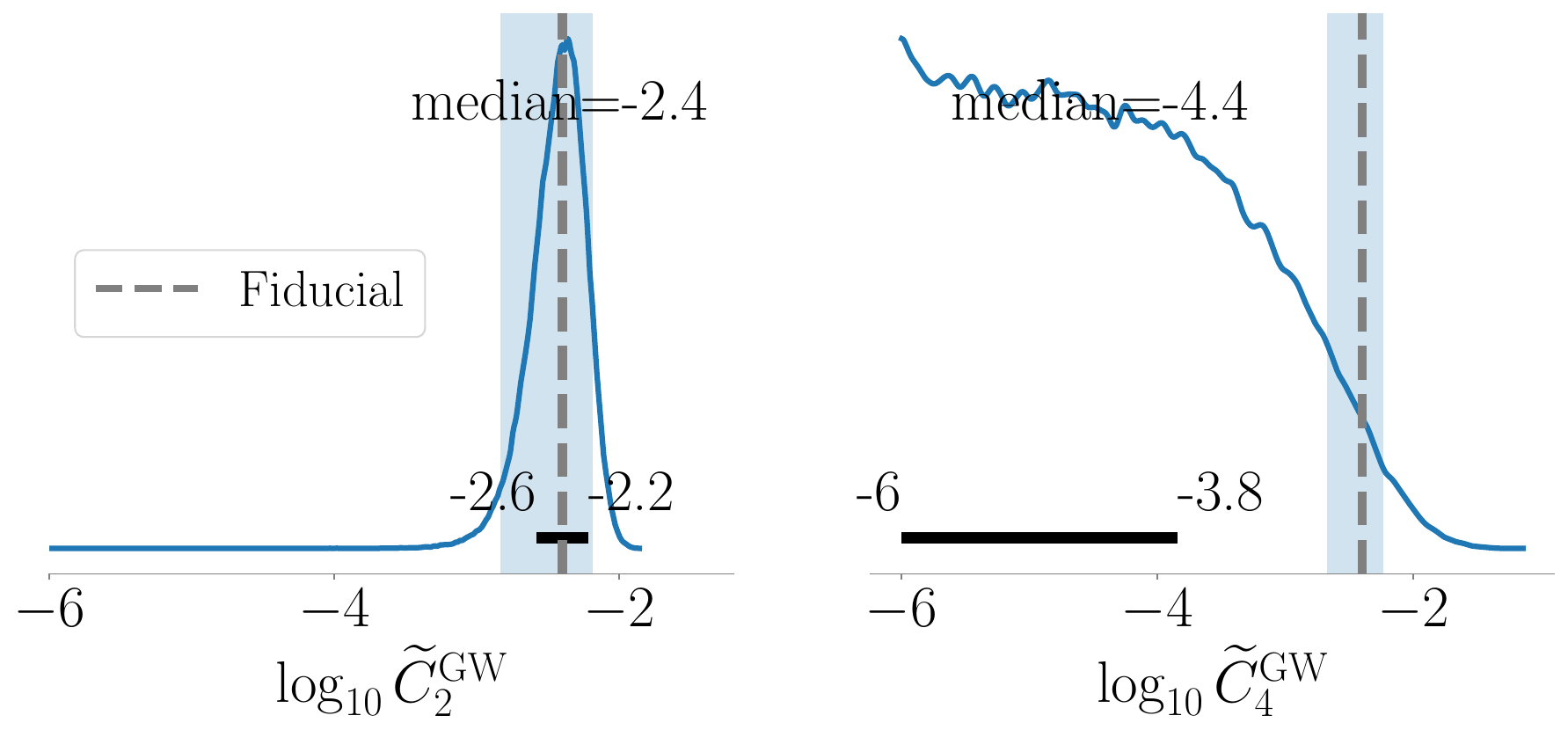}
     \includegraphics[width=0.65\linewidth]{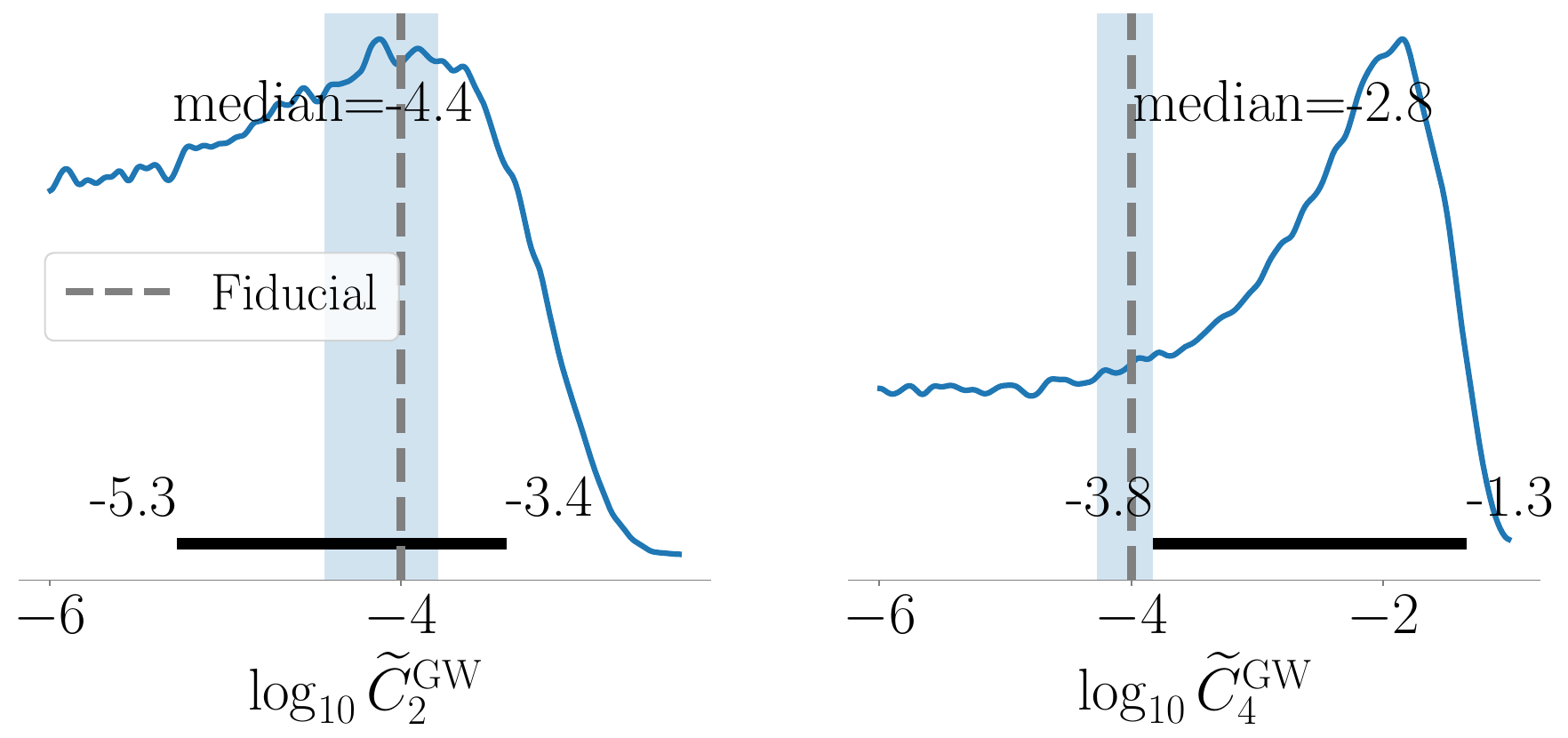}
    \caption{Reconstructed multipoles based on the LISA 4-year data,  assuming a perfect cross-correlations between the GWB and the CMB.  Upper Panel: The fiducial $\log_{10} \widetilde{C}_{\ell}^{\mathrm{GW}} = -2.4$. Lower Panel: The fiducial $\log_{10} \widetilde{C}_{\ell}^{\mathrm{GW}} = -4$. The black bars on the bottom mark the $1\sigma$ region, and the shaded regions represent the cosmic variance.}
    \label{fig:4y_cross}
\end{figure}
Depending on the sources, the anisotropies in the GWB generally show different levels of cross-correlations with diverse cosmological tracers. For the CGWB, current models suggest that it has a strong correlation with the CMB or the CMB lensing \citep{Ricciardone:2021kel, Schulze:2023ich, Ding:2023xeg}.  To explore the most optimistic constraints on the LISA detectability based on known cross-correlations, we consider an ideal scenario in which the GWB is perfectly correlated with the CMB. Under this assumption, we inject a GWB fully correlated with the CMB by setting $r^{\mathrm{GW}\times\mathrm{CMB}} = 1$,  subsequently  employing this scheme to infer the magnitude of the lowest several multipoles in the GWB. Our findings indicate that the LISA 4-year data is able to provide unbiased estimations of quadrupole $\widetilde{C}_2^{\mathrm{GW}}$ when $\log_{10} \widetilde{C}_{\ell}^{\mathrm{GW}} = -2.4$, as suggested by the upper panel in Fig. \ref{fig:4y_cross}. As the bottom panel in Fig. \ref{fig:4y_cross} implies, this reconstruction works until $\widetilde{C}_{\ell}^{\mathrm{GW}}$ decreases to approximately $\log_{10} \widetilde{C}_{\ell}^{\mathrm{GW}} \simeq -4$. 

\begin{figure}[htp]
    \centering
    \includegraphics[width=0.49\linewidth]{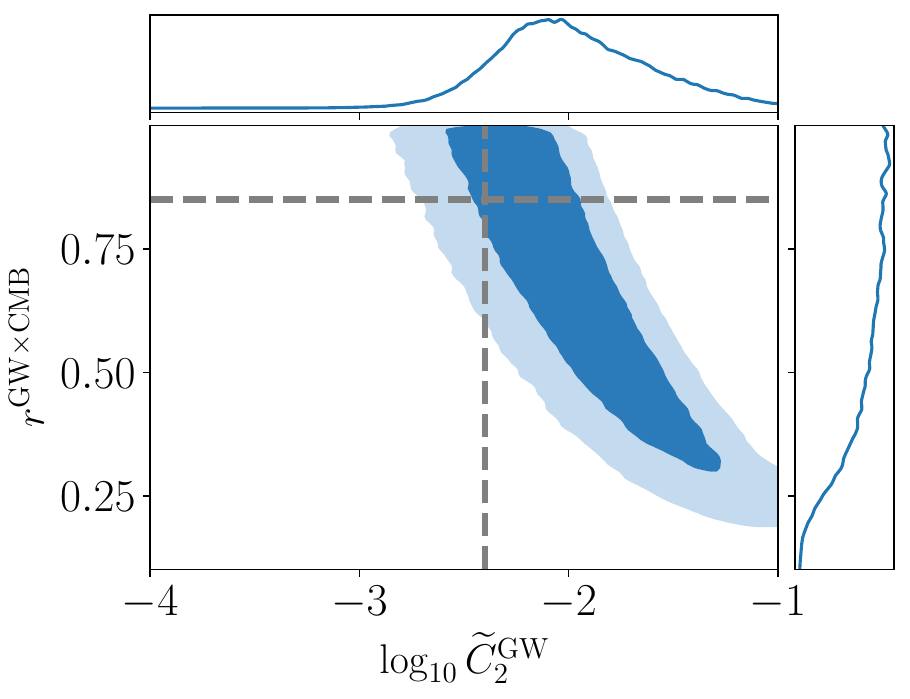}
    \includegraphics[width=0.49\linewidth]{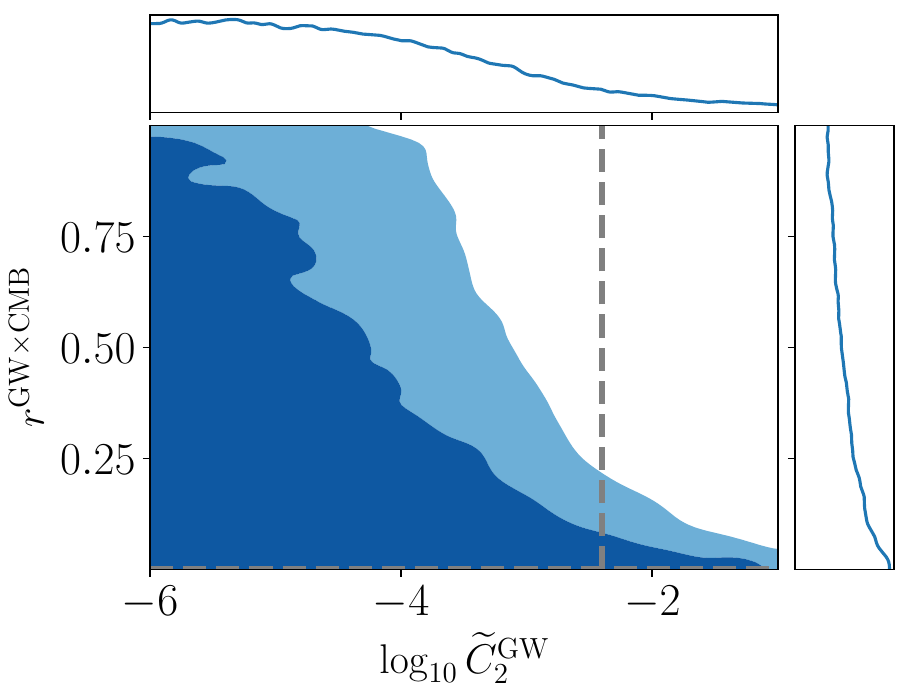}
    \caption{Contour and posterior plots of  the quadrupole $\widetilde{C}_2^{\mathrm{GW}}$ and relative cross-correlation. Shaded regions represent the $1\sigma$ and $2\sigma$ range. Left Panel: Strong correlation case: the fiducial  $r^{\mathrm{GW} \times \mathrm{CMB}}=0.85$ and $\log_{10} \widetilde{C}_{\ell}^{\mathrm{GW}}=-2.4$, denoted by the grey dashed lines. Right Panel: No correlation case: the fiducial  $r^{\mathrm{GW} \times \mathrm{CMB}}=0$ and $\log_{10} \widetilde{C}_{\ell}^{\mathrm{GW}}=-2.4$, denoted by the grey dashed lines.}
    \label{fig:4y_cross_contour}
\end{figure}

We have also simulated a GWB map based on a smaller relative correlation with the CMB, specifically $r^{\mathrm{GW}\times\mathrm{CMB}}=0.85$ with $\log_{10}\widetilde{C}_{\ell}^{\mathrm{GW}}=-2.4$. By employing the likelihood function given in eq.~\eqref{eq:likl} and taking $r^{\mathrm{GW}\times\mathrm{CMB}}$ as a new parameter, we plot  $r^{\mathrm{GW}\times\mathrm{CMB}}$ vs. $\widetilde{C}_2^{\mathrm{GW}}$ contour in the left panel of Fig.~\ref{fig:4y_cross_contour}. The results indicate that, even under the condition of a weaker correlation, our scheme remains effective in inferring quadrupole $\widetilde{C}_{2}^{\mathrm{GW}}$: the fiducial value can still be recovered within a $1\sigma$ confident level. 

Lastly, we exam the scenario where the relative correlation is minimal ($r^{\mathrm{GW}\times\mathrm{CMB}}=0.0$). As indicated in the right panel of Fig.~\ref{fig:4y_cross_contour}, due to the strong degeneracy between the strength of anisotropy and the correlation level (the smaller the anisotropy, the weaker the constraints on the cross-correlation), an upper limit for the relative correlation cannot be established. In addition, the overall constraints on the anisotropic level are similar to the case without considering the cross-correlation (see upper panel in Fig.~\ref{fig:4y_mult}). 

\section{Discussions }
In this study, we introduce a novel scheme for performing Bayesian inference on the detector data associated with  the stochastic gravitational wave background.  Utilizing a new analytical likelihood function, this approach directly connects the GW data to the angular power spectrum, with the capability to incorporate  the cross-correlations between the GWB anisotropy and any cosmological tracers. As  a crucial feature of the GWB anisotropy, these cross-correlations, predicted by various physical models, can significantly enhance the reconstruction capability of our framework.

We employ forecasted LISA response and noise to demonstrate the validity of this approach. Our findings indicate that, for the injected cosmological GWB signal, without considering any cross-correlations,  the noise level in the 4-year LISA data prevents us from constructing reliable Bayesian inference on multipoles unless the LISA mission is unrealistically extended to 80 years. However, if the anisotropies in the GWB are strongly correlated with another cosmological tracer, such as the CMB, the 4-year LISA data can provide unbiased estimates of the quadrupole moment. We emphasize that this study tests the validity of our scheme using mock data that excludes any foreground signals, which could potentially compromise the reconstruction accuracy or introduce bias into the results. Consequently, it is essential to apply appropriate foreground cleaning techniques prior to utilizing our scheme.

Bypassing the time-consuming map-making process, the posterior sampling based on this new framework is straightforward to implement, and typically completes within several hours. We also emphasize  that this scheme is designed generically for arrays of GW detectors. Our work, therefore, paves the way for detailed studies of the detectability of anisotropies in current and future GW observations, such as LVK~\citep{LIGOScientific:2016jlg}, TianQin~\citep{TianQin:2015yph}, Taiji~\citep{Hu:2017mde, Ruan:2018tsw}, DECIGO~\citep{Kawamura:2011zz}, BBO~\citep{Corbin:2005ny}, ET~\citep{Maggiore:2019uih}, and their possible joint searches~\citep{Torres-Orjuela:2023hfd, Torres-Orjuela:2024gux, Liang:2024tgn}, as well as PTA experiments. The exploration of open questions, such as how various GWB source models are constrained from the GWB anisotropy, or how to differentiate between astrophysical and cosmological sources based on the anisotropies, might benefit from this new scheme. However, the focus of this study is on the success of the method rather than the detectability for a specific detector or constraining GWB source models; we will therefore leave the detailed analysis on these topics for future studies.

\acknowledgments
Authors would like to thank Erik Floden, Vuk Mandic, Leo Tsukada and Gang Wang.  R.D. is supported in part by the National Key R\&D Program of China (No. 2021YFC2203100). C.T. is supported by the National Natural Science Foundation of China (Grants No. 12405048) and the Natural Science Foundation of Anhui Province (Grants No. 2308085QA34).   The authors acknowledge the High-performance Computing Platform of Anhui University for providing computing resources.

\appendix

\section{From the anisotropic GW fractional energy density to the GW intensity }
\label{app:a1}

In this appendix, we establish the relationship between the GW intensity multipoles $I_{\ell m}$ and the GW fractional energy multipoles $a^{\rm GW}_{\ell m}$. The metric perturbations $h_{ij}(t, \mathbf{x})$  can be expressed as a sum of contributions orginating from all directions in the sky via a plane-wave expansion:
\begin{align}
h_{i j}(t, \mathbf{x})= \int_{-\infty}^{\infty} d f \int d^2 \Omega_{\hat{\mathbf{n}}} \sum_{p =+, \times} h_p(f, \hat{\mathbf{n}}) \mathrm{e}_{i j}^p(\hat{\mathbf{n}}) e^{ i 2 \pi  f(t-\hat{\mathbf{n}} \cdot \mathbf{x})}\,,
\label{eq:1}
\end{align}
where $p=\{+, \times\}$ labels the polarization states, $\hat{\mathbf{n}}$ represents the direction of propagation in the sky. The propagation direction $\hat{\mathbf{n}}$ along with two transverse directions $\hat{\boldsymbol{\theta}}$ and $\hat{\boldsymbol{\phi}}$ are defined in terms of standard spherical coordinates:
\begin{align}
& \hat{\mathbf{n}}=\sin \theta \cos \phi \,\hat{\boldsymbol{x}}+\sin \theta \sin \phi \,\hat{\boldsymbol{y}}+\cos \theta \,\hat{\boldsymbol{z}}\,, \nonumber\\
& \hat{\boldsymbol{\theta}}=\cos \theta \cos \phi \,\hat{\boldsymbol{x}}+\cos \theta \sin \phi \,\hat{\boldsymbol{y}}-\sin \theta \, \hat{\boldsymbol{z}}\,, \nonumber\\
& \hat{\boldsymbol{\phi}}=-\sin \phi \, \hat{\boldsymbol{x}}+\cos \phi \,\hat{\boldsymbol{y}}\,.
\end{align}
With these coordinates, the GW polarization tensors are constructed as follows
\begin{align}
e_{i j}^{+}(\hat{\mathbf{n}})=\hat{\theta}_i \hat{\theta}_j-\hat{\phi}_i \hat{\phi}_j, \quad e_{i j}^{\times}(\hat{\mathbf{n}})=\hat{\theta}_i \hat{\theta}_j+\hat{\phi}_i \hat{\phi}_j\,.
\end{align}
It is easy to check that they satisfy
\begin{align}
& e_{i j}^{+}(\hat{\mathbf{n}})e^{+,ij}(\hat{\mathbf{n}})=e_{i j}^{\times}(\hat{\mathbf{n}})e^{\times, ij}(\hat{\mathbf{n}})=2\,, \nonumber\\
& e_{i j}^{+}(\hat{\mathbf{n}})e^{\times, ij}(\hat{\mathbf{n}})= e_{i j}^{\times}(\hat{\mathbf{n}})e^{+,i j}(\hat{\mathbf{n}})=0\,.
\end{align}
We thus obtain the orthogonal and normalization for the polarization tensors:
\begin{align}
e_{i j}^p(\hat{\mathbf{n}}) e^{p^\prime, i j}(\hat{\mathbf{n}}^{\prime})=2 \delta_{p p^\prime}\,.
\end{align}
Assuming that the GWB to be Gaussian, stationary, $h_{i j}$ is characterized by its first and quadratic expectation values as
\begin{align}
&\Big\langle h_p(f, \hat{\mathbf{n}})\Big\rangle=0\,, \\
&\Big\langle h_p(f, \hat{\mathbf{n}}) h_{p^{\prime}}^*\left(f^{\prime}, \hat{\mathbf{n}}^{\prime}\right)\Big\rangle \equiv \delta\left(f-f^{\prime}\right) \frac{\delta^2\left(\hat{\mathbf{n}},\hat{\mathbf{n}}^{\prime}\right)}{4 \pi} \mathrm{W}_{p p^{\prime}}(f, \hat{\mathbf{n}})\,, \label{eq:11}
\end{align}
where the matrix $W$ is related to the gravitational Stokes parameters as
\begin{align}
W(f, \hat{\mathbf{n}}) \equiv\left(\begin{array}{cc}
I(f, \hat{\mathbf{n}})+Q(f, \hat{\mathbf{n}}) & U(f, \hat{\mathbf{n}})-i V(f, \hat{\mathbf{n}}) \\
U(f, \hat{\mathbf{n}})+i V(f, \hat{\mathbf{n}}) & I(f, \hat{\mathbf{n}})-Q(f, \hat{\mathbf{n}})
\end{array}\right)\,.
\end{align}
For unpolarized GW with $Q=U=V=0$, Eq.~\ref{eq:11} is reduced to
\begin{align}
\Big\langle h_{+}(f, \hat{\mathbf{n}}) h_{+}^*\left(f^{\prime}, \hat{\mathbf{n}}^{\prime}\right)\Big\rangle = \Big\langle h_{\times}(f, \hat{\mathbf{n}}) h_{\times}^*\left(f^{\prime}, \hat{\mathbf{n}}^{\prime}\right)\Big\rangle = \delta\left(f-f^{\prime}\right) \frac{\delta^2\left(\hat{\mathbf{n}},\hat{\mathbf{n}}^{\prime}\right)}{4 \pi} I(f, \hat{\mathbf{n}})\,.
\label{eq:12}
\end{align}
Inserting Eq.~\ref{eq:1} into expression of GW energy density $\rho_{\mathrm{GW}}$ and using Eq.~\ref{eq:12}, we have
\begin{align}
\rho_{\mathrm{GW}}&=\frac{1}{32 \pi G}\Big\langle\dot{h}^{ij} \dot{h}_{ij}\Big\rangle  \nonumber\\
&= \frac{1}{32 \pi G}\int\, d f \int\, d f^\prime \int d^2 \Omega_{\hat{\mathbf{n}}} \int d^2 \Omega_{\hat{\mathbf{n}}^{\prime}} \sum_{p,p^\prime} e_{i j}^p(\hat{\mathbf{n}}) e^{p^\prime, i j}(\hat{\mathbf{n}}^{\prime}) (i 2 \pi  f)(- i 2 \pi  f^\prime) \nonumber\\
&\quad \times \Big\langle h_p(f, \hat{\mathbf{n}}) h_{p^{\prime}}^*\left(f^{\prime}, \hat{\mathbf{n}}^{\prime}\right)\Big\rangle e^{ i 2 \pi (f-f^\prime)t} e^{ i 2 \pi (f \hat{\mathbf{n}}-f^\prime \hat{\mathbf{n}}^{\prime})\cdot \mathbf{x}} \nonumber\\
&= \frac{1}{8 G}\int\,f^3 d \ln f \int d^2 \Omega_{\hat{\mathbf{n}}}\, I(f, \hat{\mathbf{n}})\,.
\end{align}
Substituting above result into the definition of GWB spectrum $\Omega_{\mathrm{GW}}(f)$ and performing a spherical-harmonic decomposition of the intensity $I(f, \hat{\mathbf{n}})$, we arrive
\begin{align}
\Omega_{\mathrm{GW}}(f) &\equiv \frac{1}{\rho_{\mathrm{c}}} \frac{d \rho_{\mathrm{GW}}}{d \ln f} \nonumber\\
&=\frac{\pi}{3 H_0^2} f^3 \int d^2 \Omega_{\hat{\mathbf{n}}}\sqrt{4\pi}\sum_{\ell=0}^{\ell_{\max }} \sum_{m=-\ell}^{\ell} I_{\ell m} (f) Y_{\ell m}(\hat{\mathbf{n}}) \nonumber\\
&= \frac{4\pi^2}{3 H_0^2} f^3 I_{0 0}(f) + \sqrt{4\pi}\frac{\pi}{3 H_0^2} f^3 \int d^2 \Omega_{\hat{\mathbf{n}}}\sum_{\ell=1}^{\ell_{\max }} \sum_{m=-\ell}^{\ell} I_{\ell m} (f) Y_{\ell m}(\hat{\mathbf{n}})\,,
\label{eq:13}
\end{align}
where we have introduced an additional $\sqrt{4\pi}$ factor to cancel $Y_{0 0}$, thereby ensuring the correspondence between the intensity monopole $I_{0 0}$ and the monopole of GW energy fraction $\overline{\Omega}_{\mathrm{GW}}$. On the other hand, we also separate the GW spectrum into the background part (monopole) and the linear fluctuation $\delta_{\rm{GW}}$, then expanding the perturbation function with spherical harmonic basis:
\begin{align}
\Omega_{\mathrm{GW}}(f) &= \frac{1}{4\pi} \int d^2 \Omega_{\hat{\mathbf{n}}} \,\overline{\Omega}_{\mathrm{GW}}(f)\left[1 + \delta_{\rm{GW}}(f, \hat{\mathbf{n}}) \right] \nonumber\\
&= \overline{\Omega}_{\mathrm{GW}}(f) + \frac{\overline{\Omega}_{\mathrm{GW}}(f)}{4\pi}\left(4-\frac{\partial \ln \overline{\Omega}_{\mathrm{GW}}\left(f \right)}{\partial \ln f}\right)\int d^2 \Omega_{\hat{\mathbf{n}}} \sum_{\ell=1}^{\ell_{\max }} \sum_{m=-\ell}^{\ell} a^{\rm GW}_{\ell m}(f) Y_{\ell m}(\hat{\mathbf{n}})\,.
\label{eq:14}
\end{align}
Equating first terms in Eqs.~\ref{eq:13} and Eq.~\ref{eq:14} give the relationship between the intensity monopole $\overline{I}_{0 0}$ and the GW monopole $\overline{\Omega}_{\mathrm{GW}}$
\begin{align}
\overline{I}_{0 0}(f) \equiv \frac{3 H_0^2}{4\pi^2}\frac{\overline{\Omega}_{\mathrm{GW}}(f)}{f^3}\,,
\end{align}
while the second terms in Eqs.~\ref{eq:13} and Eq.~\ref{eq:14} show the relationship between the GW intensity multipoles $I_{\ell m}$ and the GW energy fraction multipoles $a^{\rm GW}_{\ell m}$:
\begin{align}
I_{\ell m}(f) &= \frac{1}{\sqrt{4\pi}}\frac{3 H_0^2}{4\pi^2}\frac{\overline{\Omega}_{\mathrm{GW}}(f)}{f^3}\left(4-\frac{\partial \ln \overline{\Omega}_{\mathrm{GW}}\left(f \right)}{\partial \ln f}\right) a^{\rm GW}_{\ell m}(f)\nonumber\\
&= K_f a^{\rm GW}_{\ell m}(f)\,,
\label{eq:16}
\end{align}
where we have defined
\begin{align}
K_f \equiv {\frac{1}{\sqrt{4\pi}}\frac{3 H_0^2}{4\pi^2}\frac{\overline{\Omega}_{\mathrm{GW}}(f)}{f^3}\left(4-\frac{\partial \ln \overline{\Omega}_{\mathrm{GW}}\left(f \right)}{\partial \ln f}\right)}\,.
\label{eq:17}
\end{align}
From the auto-correlation function $\Bigl\langle a^{\rm GW}_{\ell m} a^{\rm GW *}_{\ell' m'}\Bigr\rangle= \delta_{\ell\ell'}\delta_{mm'}C^{\rm GW}_\ell$, we further obtain correspondence between the angular power spectrum of the GW intensity and  energy fraction:
\begin{align}
C^I_\ell (f) = K^2_f \,C^{\rm GW}_\ell (f)\,.
\label{eq:18}
\end{align}

\section{Conditional probability distribution considering cross-correlations}
\label{app:a2}

A multidimensional complex Gaussian distribution for $d$-dimensional random variable $\boldsymbol{x}$ is given by
\begin{align}
\mathcal{N}(\boldsymbol{\mu}, \boldsymbol{\Sigma})=\frac{1}{\pi^d |\boldsymbol{\Sigma}|}\exp \left[-(\boldsymbol{x}-\boldsymbol{\mu})^T \boldsymbol{\Sigma}^{-1}(\boldsymbol{x}-\boldsymbol{\mu})\right] \,.
\end{align}
Suppose that the vector $\boldsymbol{x} =[\boldsymbol{x}_1^T,\,\boldsymbol{x}_2^T]^T$ obeys joint gaussian distribution
\begin{align}
p(\boldsymbol{x}) &\sim \mathcal{N}\left(
\left[\begin{array}{l}
\boldsymbol{\mu}_1 \\
\boldsymbol{\mu}_2
\end{array}\right],\, \left[\begin{array}{cc}
\boldsymbol{\Sigma}_{11} & \boldsymbol{\Sigma}_{12} \\
\boldsymbol{\Sigma}_{12}^T & \boldsymbol{\Sigma}_{22}
\end{array}\right] \right) \,,
\label{eq:39}
\end{align}
where $\boldsymbol{\Sigma}_{12}$ is the non-symmetric cross-covariance matrix between $\boldsymbol{x}_1$ and $\boldsymbol{x}_2$. Then the marginal distributions are
\begin{align}
p(\boldsymbol{x}_1) =  \int p(\boldsymbol{x}) d \boldsymbol{x}_2 \sim\mathcal{N}\left(\boldsymbol{\mu}_1,\boldsymbol{\Sigma}_{11}\right), \quad p(\boldsymbol{x}_2) =  \int p(\boldsymbol{x}) d \boldsymbol{x}_1\sim\mathcal{N}\left(\boldsymbol{\mu}_2,\boldsymbol{\Sigma}_{22}\right)\,,
\end{align}
and according to the definition, the conditional  probability distributions are
\begin{align}
&p(\boldsymbol{x}_1| \boldsymbol{x}_2) \sim\mathcal{N}\left(\boldsymbol{\mu}_1+\boldsymbol{\Sigma}_{12}\boldsymbol{\Sigma}_{22}^{-1}(\boldsymbol{x}_2-\boldsymbol{\mu}_2),\boldsymbol{\Sigma}_{11}-\boldsymbol{\Sigma}_{12}\boldsymbol{\Sigma}_{22}^{-1}\boldsymbol{\Sigma}_{12}^T\right)\,, \nonumber\\
&p(\boldsymbol{x}_2| \boldsymbol{x}_1)\sim\mathcal{N}\left(\boldsymbol{\mu}_2+\boldsymbol{\Sigma}_{12}^T\boldsymbol{\Sigma}_{11}^{-1}(\boldsymbol{x}_1-\boldsymbol{\mu}_1),\boldsymbol{\Sigma}_{22}-\boldsymbol{\Sigma}_{12}^T\boldsymbol{\Sigma}_{11}^{-1}\boldsymbol{\Sigma}_{12}\right)\,.
\label{eq:31}
\end{align}
We now apply the results above to construct the likelihood function, incorporating the cross-correlations between the GWB and any known cosmological tracers. Assuming that the maps of GWB anisotropy and a set of cosmological tracers $\boldsymbol{Y}$, characterized by spherical harmonic coefficients $a^{\mathrm{GW}}_{\ell m}$ and $a^{\boldsymbol{Y}}_{\ell m}$ respectively, are multivariate Gaussian random variables with zero mean and covariance $C_{\ell}^{\mathrm{GW} \times \boldsymbol{Y}}$. Following eq.~\eqref{eq:39}, their probability distribution is given by
\begin{align}
& p\Big(a^\mathrm{GW}_{\{\ell m \}},\,a^{\boldsymbol{Y}}_{\{\ell m \}}\Big) = \prod_{\ell=1}^{\ell_{\max}} \prod_{m=-\ell}^{\ell} \frac{\exp \left\{-\left(\begin{array}{c}
a^\mathrm{GW}_{\ell m}\\
a^{\boldsymbol{Y}}_{\ell m}\end{array}\right)^T\left(\begin{array}{cc}
C_{\ell}^\mathrm{GW} & C_{\ell}^{\mathrm{GW} \times \boldsymbol{Y}} \\
C_{\ell}^{\mathrm{GW} \times \boldsymbol{Y}} & C_{\ell}^{\boldsymbol{Y}}
\end{array}\right)^{-1}\left(\begin{array}{c}
a^\mathrm{GW}_{\ell m}\\
a^{\boldsymbol{Y}}_{\ell m}
\end{array}\right)\right\}}{\pi \Big(C_{\ell}^\mathrm{GW} C_{\ell}^{\boldsymbol{Y}} -\left(C_{\ell}^{\mathrm{GW} \times \boldsymbol{Y}}\right)^2 \Big)} \,.
\end{align}
According to eq.~(\ref{eq:31}), the conditional distribution of $a^{\mathrm{GW}}_{\ell m}$ for given $a^{\boldsymbol{Y}}_{\ell m}$ obeys
\begin{align}
p\Big( a^\mathrm{GW}_{\{\ell m \}}\Big| a^{\boldsymbol{Y}}_{\{\ell m \}}, \mu^{\mathrm{GW}\mid \boldsymbol{Y}}_{\{\ell m \}}, C_{\{\ell\}}^{\mathrm{GW}\mid \boldsymbol{Y}}\Big)
= \prod_{\ell=1}^{\ell_{\max}} \prod_{m=-\ell}^{\ell} \frac{\exp \Big\{-\left(a^\mathrm{GW}_{\ell m}-\mu^{\mathrm{GW}\mid \boldsymbol{Y}}_{\ell m}\right)\left(C_{\ell}^{\mathrm{GW}\mid \boldsymbol{Y}}\right)^{-1}\left(a^\mathrm{GW}_{\ell m}-\mu^{\mathrm{GW}\mid \boldsymbol{Y}}_{\ell m}\right)\Big\}}{\pi C_{\ell}^{\mathrm{GW}\mid \boldsymbol{Y}}}  \,,
\end{align}
where the mean and the covariance are given by
\begin{align}
\mu^{\mathrm{GW}\mid \boldsymbol{Y}}_{\ell m} &= \frac{C_{\ell}^{\mathrm{GW} \times \boldsymbol{Y}}}{C_{\ell}^{\boldsymbol{Y}}} a^{\boldsymbol{Y}}_{\ell m}\,, \\
C_{\ell}^{\mathrm{GW}\mid \boldsymbol{Y}} & = C_{\ell}^{\mathrm{GW}} -\frac{\left(C_{\ell}^{\mathrm{GW} \times \boldsymbol{Y}}\right)^2}{C_{\ell}^{\boldsymbol{Y}}} = C_{\ell}^\mathrm{GW}\left[1-\left(\frac{C_{\ell}^{\mathrm{GW} \times \boldsymbol{Y}}}{\sqrt{C_{\ell}^\mathrm{GW}C_{\ell}^{\boldsymbol{Y}}}}\right)^2\right]\,.
\end{align}
From now on, for simplicity, we will focus on a single tracer, ${\mathrm{Y}}$, instead of a set of tracers, $\boldsymbol{Y}$,  although the generalization to multiple tracers is straightforward. Using eq.~(\ref{eq:Kf}), we obtain the likelihood for the GWB intensity
\begin{align}
&\calL\left(I_{\{\ell m \}, f}\mid  \mathrm{GW}\times \mathrm{Y}\right) \equiv   p\Big(I_{\{\ell m\},f} \Big| a^{{\mathrm{Y}}}_{\{\ell m \}}, \mu^{\mathrm{GW}\mid {\mathrm{Y}}}_{\{\ell m \}}, C_{\{\ell\}}^{\mathrm{GW}\mid {\mathrm{Y}}}\Big)  \nonumber\\
&=  \prod_{\ell=1}^{\ell_{\max}} \prod_{m=-\ell}^{\ell} \frac{\exp \left\{-\left(I_{\ell m}-K_f\mu^{\mathrm{GW}\mid {\mathrm{Y}}}_{\ell m}\right)\left(K^2_f C_{\ell}^{\mathrm{GW}\mid {\mathrm{Y}}}\right)^{-1}\left(I_{\ell m}-K_f\mu^{\mathrm{GW}\mid {\mathrm{Y}}}_{\ell m}\right)\right\}}{\pi C_{\ell}^{\mathrm{GW}\mid {\mathrm{Y}}}} \,.
\label{eq:58}
\end{align}

\section{Likelihood function after marginalization  }
\label{app:a3}
Utilizing the complex Gaussian integration formula
\begin{align}
\int d^{2 d}  \boldsymbol{z} \exp \left(- \boldsymbol{z}^{\dagger}  \boldsymbol{M}  \boldsymbol{z} +  \boldsymbol{K}^{\dagger}  \boldsymbol{z}+ \boldsymbol{z}^{\dagger}  \boldsymbol{K} +  \boldsymbol{Q} \right) = \frac{\pi^d}{\big| \boldsymbol{M}\big|}\exp \Big( \boldsymbol{K}^{\dagger}  \boldsymbol{M}^{-1}  \boldsymbol{K} + \boldsymbol{ Q}\Big)\,,
\label{eq:64}
\end{align}
the likelihood eq.~(\ref{eq:p_CL_D}) reads
\begin{align}
\calL\Big(D_{\{t, f\}} \Big| \mathrm{GW} \times \mathrm{Y}\Big) &= \prod_{t, f}\frac{1}{\big|\pi C_D\big|\big|C_I\big|\big|M\big|}\exp \Big(K^{\dagger}_\mu M^{-1}_{\mu\nu} K_\nu + Q \Big) \,, \label{eq:66}\\
M^{-1}_{\mu\nu} &= \left((C_I)_{\mu\nu}^{-1}+\Gamma_{\mu a}^*(C_D)_{ab}^{-1}\Gamma_{b \nu}\right)^{-1}\,, \\
K_{\mu} &= \Gamma_{\mu a}^* (C_D)_{ab}^{-1} \Big(D_b-\Gamma_{b 0} \overline{X}_0 - N_b \Big)+ (C_I)_{\mu\nu}^{-1}\overline{X}_\nu \,, \\
Q &= -\left(D_a-\overline{X}_0 \Gamma_{0 a}- N_a\right)^* (C_D)_{ab}^{-1}\left(D_b-\Gamma_{b 0} \overline{X}_0 - N_b\right)-\overline{X}_\mu^\dagger (C_I)_{\mu\nu}^{-1}\overline{X}_\nu \,.
\end{align}

We will show that eq.~(\ref{eq:66}) can be reformulated into a more concise expression, providing a clearer representation of the underlying physics. To see this, we employ Sherman-Morrison-Woodbury matrix inversion lemma, written as
\begin{align}
\left(\boldsymbol{A}+\boldsymbol{X} \boldsymbol{B} \boldsymbol{X}^\dagger\right)^{-1} &= \boldsymbol{A}^{-1}-\boldsymbol{A}^{-1} \boldsymbol{X} \left(\boldsymbol{B}^{-1}+\boldsymbol{X}^\dagger \boldsymbol{A}^{-1} \boldsymbol{X}\right)^{-1} \boldsymbol{X}^\dagger \boldsymbol{A}^{-1} \,, \\
\big|\boldsymbol{A}+\boldsymbol{X} \boldsymbol{B} \boldsymbol{X}^\dagger \big|&= \big|\boldsymbol{B}\big|\big|\boldsymbol{A}\big|\big|\boldsymbol{B}^{-1} + \boldsymbol{X}^\dagger \boldsymbol{A}^{-1} \boldsymbol{X}\big| \,,
\end{align}
where $\boldsymbol{A}$ and $\boldsymbol{B}$ are square and invertible matrices, though not necessarily have the same dimension. From which, adopting the matrix form for simplicity, we obtain following relations:
\begin{align}
\left(C_D + \Gamma C_I \Gamma^\dagger\right)^{-1} &=C_D^{-1}-C_D^{-1}\Gamma\left(C_I^{-1} + \Gamma^\dagger C_D^{-1}\Gamma\right)^{-1} \Gamma^\dagger C_D^{-1} \nonumber\\
&\simeq C_D^{-1} - C_D^{-1}\Gamma C_I \Gamma^\dagger C_D^{-1} \,, \\
\left(C_I^{-1}+ \Gamma^\dagger C_D^{-1} \Gamma\right)^{-1} &= C_I-C_I \Gamma^{+}\left(C_D+\Gamma C_I \Gamma^\dagger\right)^{-1} \Gamma C_I \nonumber\\
&\simeq C_I-C_I \Gamma^\dagger C_D^{-1} \Gamma C_I \,, \\
\big|C_D\big|\big|C_I\big|\big|M\big| &= \big|C_D\big|\big|C_I\big|\big|C_I^{-1}+ \Gamma^\dagger C_D^{-1} \Gamma\big| \nonumber\\
&= \big|C_D + \Gamma C_I \Gamma^\dagger \big| \,.
\end{align}
Following these relations, the exponential term in eq.~(\ref{eq:66}) can be rearranged to get
\begin{align}
& K^{\dagger}_\mu M^{-1}_{\mu\nu} K_\nu + Q  \nonumber\\
=& \big[\left(D_a- \Gamma_0\overline{X}_0-N_a\right)^\dagger C_D^{-1}\Gamma + \overline{X}^\dagger C_I^{-1}\big]\left(C_I^{-1} + \Gamma^\dagger C_D^{-1}\Gamma\right)^{-1} \big[\Gamma^\dagger C_D^{-1}\left(D_b- \Gamma_0\overline{X}_0 -N_b\right) + C_I^{-1}\overline{X}\big] \nonumber\\
\quad &-\left(D_a- \Gamma_0\overline{X}_0-N_a\right)^\dagger C_D^{-1} \left(D_b- \Gamma_0\overline{X}_0 -N_b\right)- \overline{X}^\dagger C_I^{-1}\overline{X} \nonumber\\
=& -\left(D_a- \Gamma_0\overline{X}_0 -N_a\right)^\dagger \big[C_D^{-1}-C_D^{-1}\Gamma\left(C_I^{-1} + \Gamma^\dagger C_D^{-1}\Gamma\right)^{-1} \Gamma^\dagger C_D^{-1}\big]\left(D_b- \Gamma_0\overline{X}_0 -N_b\right) \nonumber\\
\quad &+ \left(D_a- \Gamma_0\overline{X}_0 -N_a\right)^\dagger C_D^{-1}\Gamma \left(C_I^{-1} + \Gamma^\dagger C_D^{-1}\Gamma\right)^{-1} C_I^{-1}\overline{X} \nonumber\\
\quad &+ \overline{X}^\dagger C_I^{-1} \left(C_I^{-1} + \Gamma^\dagger C_D^{-1}\Gamma\right)^{-1}\Gamma^\dagger C_D^{-1}\left(D_b- \Gamma_0\overline{X}_0 -N_b\right) \nonumber\\
\quad &+ \overline{X}^\dagger C_I^{-1} \left(C_I^{-1} + \Gamma^\dagger C_D^{-1}\Gamma\right)^{-1} C_I^{-1}\overline{X} - \overline{X}^\dagger C_I^{-1}\overline{X}  \nonumber\\
\simeq& -\left(D_a- \Gamma_0\overline{X}_0 -N_a\right)^\dagger \left(C_D + \Gamma C_I \Gamma^\dagger\right)^{-1} \left(D_b- \Gamma_0\overline{X}_0 -N_b\right) \nonumber\\
\quad &+ \left(D_a- \Gamma_0\overline{X}_0 -N_a\right)^\dagger  \left(C_D + \Gamma C_I \Gamma^\dagger\right)^{-1} \Gamma\overline{X}\nonumber\\
\quad &+ \overline{X}^\dagger \Gamma^\dagger \left(C_D + \Gamma C_I \Gamma^\dagger\right)^{-1} \left(D_b- \Gamma_0\overline{X}_0 -N_b\right) \nonumber\\
\quad &- \overline{X}^\dagger \Gamma^\dagger \left(C_D+\Gamma C_I \Gamma^\dagger\right)^{-1} \Gamma \overline{X} \nonumber\\
=& -\left(D_a- \Gamma_0\overline{X}_0 -N_a-\Gamma\overline{X}\right)^\dagger \left(C_D + \Gamma C_I \Gamma^\dagger\right)^{-1} \left(D_b- \Gamma_0\overline{X}_0 -N_b-\Gamma\overline{X}\right)\,.
\end{align}
Taking into account the determinants, eq.~(\ref{eq:66}) then becomes
\begin{align}
&\calL\Big(D_{\{t, f\}} \Big| \mathrm{GW} \times \mathrm{Y}\Big) \nonumber \\
&= \prod_{t, f}\frac{1}{\big|\pi (C_D + \Gamma C_I \Gamma^\dagger)\big|} \exp \Big\{ -\left(D_a- \Gamma_0\overline{X}_0 -N_a-\Gamma\overline{X}\right)^\dagger \left(C_D + \Gamma C_I \Gamma^\dagger\right)^{-1} \left(D_b- \Gamma_0\overline{X}_0 -N_b-\Gamma\overline{X}\right)\Big\} \,.
\end{align}
Finally, to combine two terms associated with $\overline{X}$, we define $\overline{Z}_{\mu} \equiv \overline{X}_{\mu}$  for $\mu \geq 1$   and $\overline{Z}_{0} \equiv \overline{X}_{0}$ for $\mu= 0$, and redefine 
\begin{align}
(C_I)_{\mu\nu}\equiv K^2_f {\rm diag}\Big(\underbrace{ 0, C_{1}^{\mathrm{GW}\mid {\mathrm{Y}}}, C_{1}^{\mathrm{GW}\mid {\mathrm{Y}}}, C_{1}^{\mathrm{GW}\mid {\mathrm{Y}}},\cdots, \overbrace{C_{\ell_{\max}}^{\mathrm{GW}\mid {\mathrm{Y}}},\,\cdots,\, C_{\ell_{\max}}^{\mathrm{GW}\mid {\mathrm{Y}}}}^{(2\ell_{\max}+1)\,{\rm terms}}}_{\left(\ell_{\max}+1\right)^2\,{\rm terms}}\Big)\,,
\end{align}
to match the matrix dimensions, arriving the likelihood function in eq.~\eqref{eq:likl}
\begin{align}
&\calL\left(D_{\{t,f\}} \mid \mathrm{GW}\times \mathrm{Y} \right) \propto \prod_{t, f}{\big|\pi C_{a b}\big|^{-1}}\exp \left(-J^{\dagger}_a C^{-1}_{a b} J_b\right)\,, \label{eq:71}\\
& C_{a b} = (C_D)_{ab} + \Gamma_{a\mu} (C_I)_{\mu\nu} \Gamma_{\nu b}^* \,,\\
&J_{a} = D_{a} - \Gamma_{a \mu} \overline{Z}_{\mu} - N_{a}\,.
\end{align}

\bibliography{ref.bib} 
\bibliographystyle{JHEP} 
\end{document}